# GENERALIZED TRANSFORMS AND SPECIAL FUNCTIONS

G. DATTOLI

ENEA - Gruppo Fisica Teorica e Matematica Applicata (FISMAT)
Centro Ricerche Frascati, Rome (Italy)

E. SABIA

ENEA - Gruppo Fisica Teorica e Matematica Applicata (FISMAT)
Centro Ricerche Portici, Napoli (Italy)

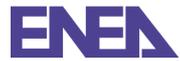

# GENERALIZED TRANSFORMS AND SPECIAL FUNCTIONS

G. DATTOLI
ENEA - Gruppo Fisica Teorica e Matematica Applicata (FISMAT)
Centro Ricerche Frascati, Rome (Italy)

E. SABIA
ENEA - Gruppo Fisica Teorica e Matematica Applicata (FISMAT)
Centro Ricerche Portici, Napoli (Italy)

# GENERALIZED TRANSFORMS AND SPECIAL FUNCTIONS

G. DATTOLI, E. SABIA

**Abstract**
We study the properties of different type of transforms by means of operational methods and discuss the relevant interplay with many families of special functions. We consider in particular the binomial transform and its generalizations. A general method, based on the use of the Fourier transform technique, is proposed for the study of the properties of functions of operators.

*Riassunto*
*In questo lavoro vengono studiate le proprietà di diversi tipi di trasformate, mediante il formalismo dei metodi operatoriali, ed è inoltre discusso il legame tra le trasformate e molte famiglie di funzioni speciali. In particolare sono trattate la trasformata binomiale e le sue generalizzazioni. Si propone un metodo generale, basato sull'uso della trasformata di Fourier, per lo studio delle proprietà delle funzioni di operatori.*

*Parole chiave: Funzioni speciali, Metodi operazionali, Trasformata integrale, Equazione integrale, Polinomi di Appèl*

# INDEX

# GENERALIZED TRANSFORMS AND SPECIAL FUNCTIONS

## 1. INTRODUCTION

In the present investigation we combine different types of transforms and discuss the relevant thread and link with the theory of special functions. We take advantage from an operational formalism, which will allow the straightforward derivation of old and new identities, within the framework of a unified point of view.

The paper consists of two parts. In the first we discuss binomial type transforms and the relevant generalizations, we show that the use of a kind of umbral notation allows a significant simplification of the associated formalism. The second part deals with the theory of functions of operators and we will show that a slightly modified form of the ordinary Fourier transform can be exploited to obtain a unifying mean, which yields a natural common frame between different types of integral transforms.

Our starting point is the so called binomial transform B. T. of a sequence $a_n$, defined as [1]

$$b_n = \sum_{s=0}^{n}(-1)^s \binom{n}{s} a_s \qquad (1).$$

The use of the umbral notation [2]

$$a_n = \hat{a}^n 1 \tag{2}$$

allows us to cast eq. (1) in the more transparent form of the Newton binomial

$$\hat{b}^n = (1 - \hat{a})^n \tag{3}.$$

Albeit apparently trivial, the previous identity allows a straightforward inversion of the B.T. and indeed we get

$$\hat{a}^n = (1 - \hat{b})^n \Rightarrow a_n = \sum_{s=0}^{n} (-1)^s \binom{n}{s} b_s \tag{4}.$$

The previous identity ensures that a B. T. coincides with its inverse, therefore, denoting a B.T. by $\hat{B}$, in such a way that

$$\hat{B}\hat{a}^n = \hat{b}^n \tag{5a}$$

we can conclude that

$$\hat{B} = \hat{B}^{-1} \Rightarrow \hat{B}^2 = \hat{1} \tag{5b}$$

and therefore transformations of the type (1) realizes an involution.

We can now specify the action of $\hat{B}$ on a series and, in particular, we will consider the series defining the following function $f(x)$[1]

$$f(x) = \sum_{n=0}^{\infty} a_n x^n \tag{6}.$$

The use of the previous notations and of eq. (3) yields

---

[1] According to our umbral notation we can write the above series given in eq. (6) as
$$f(x) = \frac{1}{1 - \hat{a}x} 1$$

$$[\hat{B}f](x) = \sum_{n=0}^{\infty} (\hat{B}a_n) x^n = \left[\sum_{n=0}^{n} (1-\hat{a})^n x^n\right] 1 = \frac{1}{1-(1-\hat{a})x} 1 \quad (7),$$

Which holds under the assumption that summation and $\hat{B}$ operators can be interchanged. This is not ensured a priori, the last identity is therefore just a formal result, since it does not specify the underlying conditions of convergence. We can, however, get a more rigorous and meaningful result, by noting that

$$\frac{1}{1-(1-\hat{a})x} 1 = \frac{1}{1-x} \frac{1}{1+\frac{\hat{a}x}{1-x}} 1 = \frac{1}{1-x} \sum_{s=0}^{\infty} \frac{(-x)^n}{(1-x)^n} \hat{a}^n 1 = \frac{1}{1-x} f(-\frac{x}{1-x}),$$

$$|x| < 1 \quad (8),$$

Accordingly we obtain

$$[\hat{B}f](x) = \frac{1}{1-x} f(-\frac{x}{1-x})$$

$$|x| < 1 \quad (9).$$

The action of $\hat{B}$ on exponential like functions of the type $g(x) = \sum_{n=0}^{\infty} \frac{a_n}{n!} x^n$ [2], can be understood by the use of the same procedure leading to eq. (9), which yields

$$[\hat{B}g](x) = \left[\sum_{n=0}^{\infty} \frac{(1-\hat{a})^n}{n!} x^n\right] 1 = e^x g(-x) \quad (10).$$

The so far obtained results are well known in the theory of B. T., but have been obtained here by means of a straightforward procedure, which will be shown to be very flexible and amenable for further generalizations.

---

[2] By exponential like we mean that, according to the umbral notation, we can write $g(x) = e^{(\hat{a}x)} 1$

## 2. THE MODULAR TRANSFORM AND OTHER GENERALIZATIONS

This section is devoted to the extension of the previous formalism to some generalization of the binomial transform proposed in the past.

Prodinger has defined the following modular transform $\hat{B}(\alpha,\beta)$ on a sequence $a_n$ as the following generalization of the B. T. [3]

$$b_n = \sum_{s=0}^{n} (-1)^s \binom{n}{s} \alpha^{n-s} \beta^s a_s \qquad (11).$$

According to the previous suggestions, we can be write eq. (11) in the form

$$\hat{b}^n = (\alpha - \beta \hat{a})^n \qquad (12)$$

and therefore, in analogy to what has already been done, we find

$$\left[\hat{B}(\alpha,\beta) f\right](x) = \frac{1}{1-\alpha x} f\left(\frac{\beta x}{\alpha x - 1}\right),$$

$$|x| < \alpha^{-1}, \qquad (13)$$

$$\left[\hat{B}(\alpha,\beta) g\right](x) = e^{\alpha x} g(-\beta x)$$

The inversion of the modular transform is therefore easily obtained as

$$a_n = \frac{1}{\beta^n} \sum_{s=0}^{n} (-1)^s \binom{n}{s} \alpha^{n-s} b_s \qquad (14).$$

and therefore $\hat{B}^{-1}(\alpha,\beta) = \frac{1}{\beta^n} \hat{B}(\alpha,1)$.

The previously obtained results can also be exploited to study the rising k-binomial transform [4] $^{(k)}\hat{B}$, usually defined through the relation

$$b_n = \sum_{s=0}^{n} (-1)^s \binom{n}{s} s^k a_s \tag{15},$$

which can be written in umbral form after setting

$$s^k = (t\partial_t)^k t^s \mid_{t=1} \tag{16}.$$

It is therefore evident the umbral counterpart of eq. (15) is

$$\hat{b}^n = (t\partial_t)^k (1 - t\hat{a})^n \mid_{t=1} \tag{17}.$$

The use of the previous results yields therefore

$$\left[^{(k)}Bf\right](x) = (t\partial_t)^k \left[\hat{B}(1,t)f\right](x) = \frac{1}{1-x}(t\partial_t)^k f(-\frac{tx}{1-x}) \tag{18}$$

and the identity [5]

$$(t\partial_t)^n = \sum_{k=0}^{n} S_2(k,n) t^k \partial_t^k \tag{19}$$

with $S_2(k,n)$ being Stirling number of the second kind defined as

$$S_2(k,n) = \frac{1}{k!} \sum_{j=0}^{k} (-1)^{k-j} \binom{k}{j} j^n \tag{20},$$

eventually provides us with

$$\left[^{(k)}Bf\right](x) = \sum_{r=0}^{k} \frac{(-x)^r}{(1-x)^{r+1}} S_2(r,k) f^{(r)}(-\frac{x}{1-x}),$$
$$|x| < 1 \tag{21}.$$

Where we have denoted by $f^{(r)}(.)$ the r-th derivative of the $f$ function. Furthermore the action on the exponential like functions is easily obtained in the form

$$\left[{}^{(k)}\hat{B}g\right](x) = e^x \sum_{r=0}^{k} (-x)^r S_2(r,k) g^{(r)}(-x) \tag{22}$$

This section has perhaps given a feeling of the usefulness of the umbral formalism for the study of binomial type transforms, in the following section we will discuss further generalizations.

## 3. THE HERMITE AND LAGUERRE TRANSFORMS

We define the Hermite transform of a sequence $a_n$ as

$$b_n = \sum_{r=0}^{[n/2]} \binom{n}{r}_2 \alpha^{n-2r} \beta^r a_r,$$

$$\binom{n}{r}_2 = \frac{n!}{(n-2r)!r!} \tag{23}$$

and on account of the fact that

$$H_n(x,y) = n! \sum_{r=0}^{[n/2]} \frac{x^{n-2r} y^r}{(n-2r)!r!} \tag{24}$$

are two variable Hermite polynomials [6], satisfying the generating function

$$\sum_{n=0}^{\infty} \frac{t^n}{n!} H_n(x,y) = e^{xt+yt^2} \tag{25}$$

we can write the umbral counterpart of eq. (22) as

13$$\hat{b}^n = H_n(\alpha, \beta\,\hat{a}) \tag{26}$$

The action of such a transform on the series $g(x)$ can be written as follows

$$\left[\hat{H}(\alpha,\beta)g\right](x) = \sum_{n=0}^{\infty} \frac{1}{n!} H_n(\alpha,\beta\,\hat{a})\, x^n 1 = e^{\alpha x + \beta\,\hat{a}\, x^2} 1 = e^{\alpha x} g(\beta x^2) \tag{27}$$

On the other side the complementary Hermite transform

$$b_n = \sum_{r=0}^{[n/2]} \binom{n}{r}_2 \alpha^{n-2r} \beta^r a_{n-2r}, \Rightarrow \hat{b}^n = H(\alpha\,\hat{a},\beta) \tag{28}$$

yields

$$\left[\hat{\tilde{H}}(\alpha,\beta)g\right](x) = \sum_{n=0}^{\infty} \frac{1}{n!} H_n(\alpha\,\hat{a},\beta)\, x^n 1 = e^{\alpha\,\hat{a}\, x + \beta\, x^2} 1 = e^{\beta x^2} g(\alpha x) \tag{29}$$

We can now discuss the possibility of defining the composition of different transforms and indeed by the notation

$$\hat{C}(\gamma,\delta;\alpha,\beta) = \hat{\tilde{H}}(\gamma,\delta) \circ \hat{B}(\alpha,\beta) \tag{30}$$

we will just indicate a transform consisting of the successive application of binomial and Hermite transforms. By recalling the "addition theorem" [7]

$$\sum_{s=0}^{n} \binom{n}{s} x^s H_{n-s}(y,z) = H_n(x+y,z) \tag{31}$$

it easily follows that eq. (29) is equivalent to the transform

$$\hat{b}^n = H_s(-\beta\gamma\,\hat{a} + \alpha, \beta^2\delta) \tag{32}$$

Analogous conclusions can be drawn for compositions involving the repeated action of Hermite transforms, as we will discuss in the concluding section[3].

A further example of generalization of B. T. the Laguerre transform, defined as

$$b_n = \sum_{r=0}^{n} l_{n,r} \beta^{n-r} \alpha^r a_r,$$

$$l_{n,r} = \frac{(-1)^r}{(r!)^2 (n-r)!}$$

(33).

In umbral terms it can be written as

$$\hat{b}^n = L_n(\alpha \hat{a}, \beta) \tag{34}$$

where $L_n(x,y)$ denotes two variable Laguerre polynomials [8].

The action of the Laguerre transform on $f(x)$ yields

$$\left[\hat{L}(\alpha,\beta) f\right](x) = \frac{1}{1-\beta x} g\left(-\frac{\alpha x}{1-\beta x}\right),$$

$$\left[\hat{L}(\alpha,\beta) g\right](x) = e^{\beta x} q(-x),$$

$$q(x) = \sum_{r=0}^{\infty} a_r \frac{x^r}{(r!)^2}$$

(35)

as noticeable examples we stress that for $g(x) = e^x$ we get

$$\left[\hat{L}(\alpha,\beta) g\right](x) = e^{\beta x} J_0(2\sqrt{x}), \tag{36}$$

---

3   The inversion of the Hermite transform is ensured by the following identities

$z^n = H_n(\alpha x, y),$

$x^n = \alpha^{-n} H_n(z, -y)$

implying that $\hat{a}^n = \alpha^{-n} H_n(\hat{b}, -\beta)$

while for $f(x) = \dfrac{1}{1-x}$ the first of eqs. (35) gives

$$[\hat{L}(\alpha,\beta)f](x) = \dfrac{1}{1-\beta x} e^{-\dfrac{\alpha x}{1-\beta x}}, \tag{37}$$

The results obtained in this section complete the first part of the paper and we have shown that very general results in the theory of binomial transforms can be obtained using a point of view which is essentially a restatement of the theory of generating function.

## 4. EXPONENTIAL OPERATORS AND THE FOURIER TRANSFORM

In the first part of the paper we have shown how the combined use of operational methods and special functions is a sufficiently powerful tool to deal with many aspects of binomial transform and of its generalization. In this section we consider the extension of the methods to the theory of exponential operators and Fourier transform (F.T.)

The following example will provide a first hint of the procedure, we will follow and eventually generalize.

As is well known the use of F.T. methods simplifies the evaluation of many calculations involving the action of an exponential operator on a given function, as it happens in the solution of heat type equations, like

$$\begin{aligned}\partial_\alpha F(x,\alpha) &= \partial_x^2 F(x,\alpha), \\ F(x,0) &= f(x)\end{aligned} \tag{38}$$

Whose solution can formally be written as

$$F(x,\alpha) = e^{\alpha \partial_x^2} f(x) \tag{39}$$

If we make the assumption that $f(x)$ admits an F.T. we can write

$$e^{\alpha \partial_x^2} f(x) = e^{\alpha \partial_x^2} \frac{1}{\sqrt{2\pi}} \int_{-\infty}^{+\infty} \tilde{f}(k) e^{ikx} dk = \frac{1}{\sqrt{2\pi}} \int_{-\infty}^{+\infty} \tilde{f}(k) e^{-\alpha k^2} e^{ikx} dk,$$

$$\tilde{f}(k) = \frac{1}{\sqrt{2\pi}} \int_{-\infty}^{+\infty} f(x) e^{-ikx} dx$$

(40)

According to the above procedure the problem of getting the solution of eq. (38) consists in finding the anti-F. T. of the function $\tilde{f}(k) e^{-k^2}$.

We underline that the method works only if the function $f(x)$ admits an F. T.. If not we should follow a less "natural" method.

Let us therefore consider the problem of evaluating the following expression

$$F(x) = \Phi(\partial_x) f(x) \qquad (41)$$

with $f(x)$ not admitting any F.T. . If, on the other side, the function $\Phi(x)$ possesses an F. T., we can write

$$\Phi(\partial_x) = \frac{1}{\sqrt{2\pi}} \int_{-\infty}^{+\infty} \tilde{\Phi}(k) e^{ik\partial_x} dk \qquad (42)$$

The use of the identity $e^{\lambda \partial_x} f(x) = f(x + \lambda)$ yields

$$F(x) = \frac{1}{\sqrt{2\pi}} \int_{-\infty}^{+\infty} \tilde{\Phi}(k) f(x + ik) dk \qquad (43).$$

Equation (43) realizes a new kind of transform, having interesting properties, which will be discussed below.

Just to give a simple, but important example, we consider the solution of the following initial value problem

$$\partial_y F(x,y) = -\partial_x^2 F(x,y),$$

$$F(x,0) = x^n \tag{44}$$

The initial function $f(x) = x^n$ does not admit an F. T., we write therefore the relevant solution in the form

$$F(x,y) = \Phi(\partial_x) f(x)$$

$$\Phi(x) = e^{-y x^2}, \tag{45}$$

$$f(x) = x^n$$

The solution of eq. (44) is therefore easily derived in the form

$$F(x,y) = \frac{1}{2\sqrt{\pi y}} \int_{-\infty}^{\infty} e^{-\frac{k^2}{4y}} (x+ik)^n dk \tag{46},$$

which, according to the identity [8]

$$H_n(x,-y) = e^{-y \partial_x^2} x^n \tag{47},$$

can also be viewed as an integral representation of the two variable Hermite polynomials. The use of the identity $x^n = e^{-y\partial_x^2} H_n(x,y)$ yields the further relation

$$x^n = \frac{1}{2\sqrt{\pi y}} \int_{-\infty}^{\infty} e^{-\frac{k^2}{4y}} H_n(x+ik,y) dk \tag{48}.$$

which expresses the ordinary monomials in terms of an integral representation involving Hermite polynomials.

It is evident hat the method applies to operator function whose argument is an operator more complicated then a simple derivative

$$\Phi(\alpha \partial_x + \beta x) = \frac{1}{\sqrt{2\pi}} \int_{-\infty}^{+\infty} \tilde{\Phi}(k) e^{i k (\alpha \partial_x + \beta x)} dk \tag{49}$$

If we disentangle the exponential using the Weyl decoupling rule[4], we eventually end up with

$$\Phi(\alpha \partial_x + \beta x) f(x) = \frac{1}{\sqrt{2\pi}} \int_{-\infty}^{+\infty} \tilde{\Phi}(k) e^{-\frac{\alpha \beta}{2} k^2 + i k \beta x} f(x + i\alpha k) dk \tag{50}.$$

The integral transform on the rhs of eq. (50) can be viewed as a generalization of the Gabor transform [9].

Let us now consider the following integral equation

$$\partial_\tau F(x,\tau) = -\int_0^x dx' \int_0^{x'} dx'' F(x'',\tau), \tag{51}$$

$$F(x,0) = f(x)$$

Which can also be written as

$$\partial_\tau F(x,\tau) = -\hat{D}_x^{-2} F(x,\tau) \tag{52}$$

Where $\hat{D}_x^{-1}$ is a negative derivative operator [8], whose n-th action on a given function can be defined in terms of the Cauchy repeated integral

$$\hat{D}_x^{-n} f(x) = \frac{1}{\Gamma(n)} \int_0^x (x - \xi)^{n-1} f(\xi) d\xi \tag{53}.$$

According to our previous suggestions, the solution of eq. (52) can be written as

$$F(x,\tau) = \frac{1}{2\sqrt{\pi\tau}} \int_{-\infty}^{\infty} e^{-\frac{k^2}{4\tau}} e^{ik\hat{D}_x^{-1}} f(x) dk \quad (54).$$

---

[4] By Weyl rule we mean the disentanglement identity according to which the exponential operator raised to the sum of two non commuting operators can be cast as an ordered product of exponentials, namely: $e^{\hat{A}+\hat{B}} = e^{\hat{A}} e^{\hat{B}} e^{-\frac{k}{2}}$ $k = [\hat{A},\hat{B}]$ if $k$ is a c-number or any operator commuting with the operators $\hat{A}$ and $\hat{B}$.

If for example $f(x) = 1$, the above solution can explicitly be written as

$$F(x,\tau) = \frac{1}{2\sqrt{\pi\tau}} \int_{-\infty}^{\infty} e^{-\frac{k^2}{4\tau}} C_0(-ikx) dk \quad (55)$$

where $C_0(x)$ is the 0-th order Tricomi- Bessel function.

The proof of eq. (55) can easily be achieved by noting that

$$e^{-\alpha \hat{D}_x^{-1}} 1 = \sum_{r=0}^{\infty} \frac{(-\alpha)^r \hat{D}_x^{-r}}{r!} 1 = \sum_{r=0}^{\infty} \frac{(-\alpha)^r x^r}{r!^2} = C_0(x) \quad (56).$$

On the other side if $f(x)$ is an exponential like function, we can write our solution in the form

$$F(x,\tau) = \frac{1}{2\sqrt{\pi\tau}} \sum_{n=0}^{\infty} a_n x^n \int_{-\infty}^{\infty} e^{-\frac{k^2}{4\tau}} C_n(-ikx) dk,$$

$$C_n(x) = \sum_{r=0}^{\infty} \frac{(-1)^r x^r}{r!(n+r)!} \quad (57),$$

as a consequence of the fact that

$$e^{-\alpha \hat{D}_x^{-1}} f(x) = \sum_{n=0}^{\infty} \frac{a_n}{n!} \sum_{r=0}^{\infty} \frac{(-\alpha)^r \hat{D}_x^{-r}}{r!} x^n =$$
$$= \sum_{n=0}^{\infty} \frac{a_n}{n!} \sum_{r=0}^{\infty} \frac{(-\alpha)^r}{r!(n+r)!} x^{n+r} = \sum_{n=0}^{\infty} \frac{a_n}{n!} x^n C_n(x) \quad (58).$$

In the following section we will explore the importance of the previous results within the framework of the theory of the special polynomials.

## 5. THE APPÈL POLYNOMIALS AND THE FOURIER TRANSFORM

In this section we will show how the combined use of the F. T. methods and of operational techniques yields the possibility of handling the theory of the Appèl polynomials (A. P.) using a very simple and effective formalism which allows a new method for the expansion of functions in terms of A. P., even though they provide orthogonal families in a restricted number of cases only.

The A. P. are defined by means of the generating function [10]

$$\sum_{n=0}^{\infty} \frac{t^n}{n!} a_n^+(x) = A(t) e^{tx} \qquad (59a)$$

and according to the point of view developed in ref. [11] their complementary counterparts can be defined as

$$\sum_{n=0}^{\infty} \frac{t^n}{n!} a_n^-(x) = [A(t)]^{-1} e^{tx} \qquad (59b).$$

the function $A(t)$ (as well as $[A(t)]^{-1}$), appearing in eqs. (59), is assumed to be differentiable, at least once, and free from singularities.

An alternative operational definition just following from eqs. (59), is given below

$$a_n^+(x) = A(\partial_x) x^n,$$
$$a_n^-(x) = [A(\partial_x)]^{-1} x^n \qquad (60)$$

In this letter we take advantage from the above definitions to establish a general technique to expand a given function in terms of "+" or "−" Appèl polynomials.

We consider the formal expansion of a function $f(x)$ in terms of "+" polynomials, by writing

$$f(x) = \sum_{n=0}^{\infty} \alpha_n^+ a_n^+(x) \qquad (61).$$

The coefficients $\alpha_n^+$ of the expansion can be evaluated, by following a procedure, which is based on the properties expressed by the operational identities reported in eqs. (60). According to which we find

$$[A(\partial_x)]^{-1} f(x) = \sum_{n=0}^{\infty} \alpha_n^+ x^n \qquad (62)$$

By making the assumption that the function $f(x)$ admits a Fourier transform we can rewrite eq. (62) as

$$\frac{1}{\sqrt{2\pi}} \int_{-\infty}^{\infty} \tilde{f}(k)[A(\partial_x)]^{-1} e^{ikx} dk = \frac{1}{\sqrt{2\pi}} \int_{-\infty}^{\infty} \tilde{f}(k)[A(ik)]^{-1} e^{ikx} dk =$$
$$= \sum_{n=0}^{\infty} \alpha_n^+ x^n \qquad (63).$$

The coefficient of the expansion can now be obtained by expanding the exponential, under the integral sign, and by comparing the like $x$ power terms, thus getting

$$\alpha_n^+ = \frac{i^n}{\sqrt{2\pi}\, n!} \int_{-\infty}^{\infty} \tilde{f}(k)[A(ik)]^{-1} k^n dk \qquad (64).$$

Before proceeding with some examples, we note that the above expansion holds even though not all the A. P. form an orthogonal family.

The Bernoulli polynomials belong to the Appèl family, and the relevant characteristic function $A(t)$ writes

$$[A(t)]^{-1} = \frac{e^t - 1}{t} \qquad (65)$$

the expansion coefficients for a Gaussian write therefore

$$\alpha_n^+ = \frac{i^{n-1}}{\sqrt{2\pi}\, n!} \int_{-\infty}^{\infty} e^{-\frac{k^2}{4}} (e^{ik} - 1) k^{n-1} dk \tag{66}$$

In a forthcoming investigation we will discuss in greater detail the method of A. P. expansion, here we note that it is just a by-product of the generalized transform method we have developed in the paper.

## 6. CONCLUDING REMARKS

The methods we have considered so far applies to a large variety of problems. Let us for example consider the following product of operators

$$\hat{P} = f(\partial_x) g(x \partial_x) \tag{67}$$

Whose handling may be rather cumbersome, if ordinary tools are to be employed. On the other side if both functions, on the r. h. s. of eq. (67) have an F. T., we get

$$\hat{P} = \frac{1}{2\pi} \int_{-\infty}^{\infty} \tilde{f}(k) e^{ik\partial_x} dk \int_{-\infty}^{\infty} \tilde{g}(\eta) e^{i\eta x \partial_x} d\eta =$$

$$= \frac{1}{2\pi} \int_{-\infty}^{\infty} dk\, \tilde{f}(k) \left[ \int_{-\infty}^{\infty} d\eta\, \tilde{g}(\eta) e^{i\eta x \partial_x} e^{-k(\eta - i)\partial_x} \right] \tag{68}$$

Thus finding

$$\hat{P} F(x) = \frac{1}{2\pi} \int_{-\infty}^{\infty} dk\, \tilde{f}(k) \left[ \int_{-\infty}^{\infty} d\eta\, \tilde{g}(\eta) F(e^{i\eta} x - k\eta + ik) \right] \tag{69}$$

We have already stressed that the combined use of the methods proposed here and of standard disentanglement procedures, usually adopted to deal with exponential operators, may allow the solution of problems hardly achievable with other means. An interesting example is provided indeed by the operator

$$\hat{O} = f(\alpha \partial_x^2 + \beta x) = \frac{1}{\sqrt{2\pi}} \int_{-\infty}^{\infty} \tilde{f}(k) e^{ik(\alpha \partial_x^2 + \beta x)} dk \tag{70}$$

The problem of defining a convenient ordered form for the exponential operator appearing in the r. h. s. of eq. (70) is easily achieved by setting

$$\hat{A} = ik\alpha \partial_x^2,$$
$$\hat{B} = ik\beta x \tag{71}$$

and by noting that

$$[\hat{A}, \hat{B}] = m \hat{A}^{\frac{1}{2}},$$
$$m = 2(ik)^{\frac{3}{2}} \alpha^{\frac{1}{2}} \beta \tag{72}.$$

According to eq. (72), we can apply the disentanglement rule [8]

$$e^{\hat{A}+\hat{B}} = e^{\frac{m^2}{12} - \frac{m}{2}\hat{A}^{\frac{1}{2}} + \hat{A}} e^{\hat{B}} \tag{73}$$

Thus finding

$$\hat{O} = \frac{1}{\sqrt{2\pi}} \int_{-\infty}^{\infty} \tilde{f}(k) e^{\hat{\Phi}} e^{ik\beta x} dk,$$
$$\hat{\Phi} = -i\frac{k^2}{3}\alpha \beta^2 - k^2 \alpha \beta \partial_x + ik\alpha \partial_x^2 \tag{74}$$

The action of the operator $\hat{O}$ on the monomial $x^n$ yields the interesting expression

$$\hat{O} x^n = \frac{1}{\sqrt{2\pi}} \int_{-\infty}^{\infty} \tilde{f}(k) e^{-\frac{10}{3} i k^3 \alpha \beta^2} e^{ik\beta x} H_n(x - 2k^2 \beta \alpha, ik\alpha) dk \tag{75}.$$

An other important example is offered by the possibility of evaluating operator functions containing generators of angular momentum operators, as e. g.

$$\hat{I} = f(\Omega \hat{\sigma}_+ + \Omega^* \hat{\sigma}_-),$$
$$[\hat{\sigma}_+, \hat{\sigma}_-] = 2\hat{\sigma}_3 \tag{76}$$

Where the $\hat{\sigma}$ are the Pauli matrices specified below

$$\hat{\sigma}_+ = \begin{pmatrix} 0 & 1 \\ 0 & 0 \end{pmatrix}, \hat{\sigma}_- = \begin{pmatrix} 0 & 0 \\ 1 & 0 \end{pmatrix},$$
$$\hat{\sigma}_3 = \frac{1}{2}\begin{pmatrix} 1 & 0 \\ 0 & -1 \end{pmatrix} \tag{77}$$

If $\Omega = i|\Omega|$ and the $f$ functions appearing in eq. (76) admits an F. T., we can write

$$\hat{I} = \frac{1}{\sqrt{2\pi}} \int_{-\infty}^{\infty} \tilde{f}(k) e^{-\hat{i}|\Omega|k} dk,$$
$$\hat{i} = \begin{pmatrix} 0 & 1 \\ -1 & 0 \end{pmatrix} \tag{78}$$

By recalling that

$$e^{\hat{i}\alpha} = \begin{pmatrix} \cos(\alpha) & \sin(\alpha) \\ -\sin(\alpha) & \cos(\alpha) \end{pmatrix} \tag{79}$$

we obtain from eq. (78)

$$\hat{I} = \frac{1}{\sqrt{2\pi}} \int_{-\infty}^{\infty} \tilde{f}(k) \begin{pmatrix} \cos(|\Omega|k) & -\sin(|\Omega|k) \\ \sin(|\Omega|k) & \cos(|\Omega|k) \end{pmatrix} dk, \tag{80}.$$

In a forthcoming investigation we will more carefully discuss the impact of the method on the functions of matrices, along with more specific applications for the solution of quantum mechanical problems.

As a further example, we will consider the solution of the following initial value problem involving an integro-differential operator[5]

$$\partial_\tau F(x,\tau) = -\left[\partial_x x \partial_x + \beta \hat{D}_x^{-1}\right]^m F(x,\tau)$$
$$F(x,0) = f(x)$$
(81).

The solution of the previous equation is rather complicated for a conventional procedure. On the other side the technique developed so far yields the solution in the following straightforward form

$$F(x,\tau) = \frac{1}{\sqrt{2\pi}} \int_{-\infty}^{\infty} \tilde{e}_m(k,\tau) e^{ik(\partial_x x \partial_x + \beta \hat{D}_x^{-1})} f(x) dk,$$

$$e_m(k,\tau) = \frac{1}{\sqrt{2\pi}} \int_{-\infty}^{\infty} e^{-\tau x^m - ikx} dx$$
(82).

The operator $_L\hat{D}_x = \partial_x x \partial_x$ is the so called Laguerre derivative [8, 12] and it is easily understood that the following commutation relation holds[6]

$$\left[_L\hat{D}_x, \hat{D}_x^{-1}\right] = 1$$
(83).

The Laguerre derivative and the integral operator realize a Weyl algebra, we can therefore apply the already exploited disentanglement procedure to get

---

[5] If for example $m=2$, eq. (81) explicitly reads
$$\partial_\tau F(x,\tau) = -\left[(\partial_x x \partial_x)^2 F(x,\tau) + \beta^2 \int_0^x dx' \int_0^{x'} dx'' F(x'',\tau) + 2x \partial_x F(x,\tau) + F(x,\tau)\right]$$

[6] We remind that the operator $\hat{D}_x^{-1}$ is defined in such a way that $\hat{D}_x^{-1} f(x) = \int_0^x f(x') dx'$, the commutation relation (83) holds only if the operators $_L\hat{D}_x, \hat{D}_x^{-1}$ act on functions such that $f(0)=0$.

$$F(x,\tau) = \frac{1}{\sqrt{2\pi}} \int_{-\infty}^{\infty} \tilde{e}_m(k,\tau) e^{-\frac{\beta k^2}{2}} e^{i\beta k \hat{D}_x^{-1}} e^{ik\,\partial_x x \partial_x} f(x)\, dk \qquad (84).$$

The action of the exponential operator raised to $_L\hat{D}_x$ on a given function $f(x)$ can be specified as it follows

$$e^{\alpha\,\partial_x x \partial_x} f(x) = f_B(\hat{D}_x^{-1} + \alpha),$$
$$f_B(x) = \int_0^{\infty} f(sx) e^{-s}\, ds \qquad (85)$$

With $f_B(x)$ being the Borel transform of the function $f(x)$.

If for example $f(x) = C_0(x)$ we find $f_B(x) = e^{-x}$, thus getting from eq. (82)

$$F(x,\tau) = \frac{1}{\sqrt{2\pi}} \int_{-\infty}^{\infty} \tilde{e}_m(k,\tau) e^{-\frac{\beta k^2}{2}} e^{ik + i\beta k \hat{D}_x^{-1}}\, dk\, e^{\hat{D}_x^{-1}} =$$
$$= \frac{1}{\sqrt{2\pi}} \int_{-\infty}^{\infty} \tilde{e}_m(k,\tau) e^{-\frac{\beta k^2}{2} + ik} C_0(x + i\beta k)\, dk \qquad (86).$$

In the first part of the paper we have introduced, among the other things, the Hermite transform, which can be reformulated using the technique outlined in the second part.

According to the umbral notation we can write use the same rule of the ordinary calculus and introduce, along with the operator $\hat{a}$, its derivative $\partial_{\hat{a}}$ such that

$$\partial_{\hat{a}} \hat{a}^n = n\, \hat{a}^{n-1} \qquad (87)$$

In this way we find

$$e^{y \partial_{\hat{a}}^2} g(x) = e^{y \partial_{\hat{a}}^2} e^{\hat{a} x} = e^{y x^2} e^{\hat{a} x} = e^{y x^2} g(x) \qquad (88)$$

Which is the same result reported in eq. (29).

More complicate transforms can be handled too, and indeed we find

$$F(\partial_{\hat{a}})f(x) = \frac{1}{\sqrt{2\pi}} \int_{-\infty}^{\infty} \tilde{F}(k) e^{ik\partial_{\hat{a}}} \sum_{n=0}^{\infty} (\hat{a}x)^n \, dk =$$

$$= \frac{1}{\sqrt{2\pi}} \int_{-\infty}^{\infty} \frac{\tilde{F}(k)}{1+ikx} f\left(\frac{x}{1+ikx}\right) dk \qquad (89).$$

In a forthcoming investigation we will explore more carefully the points just touched on in these concluding remarks and some applications to problems in quantum mechanics.

**ACKNOWLEDGEMENTS**